\documentclass[12pt]{iopart}
\usepackage{iopams}  
\usepackage{graphicx}

\def\fig#1{Fig.~\ref{#1}}

\newcommand{\ptt} {\mbox{$p_T$}}
\newcommand{\pttrig} {\mbox{$p_{Ttrig}$}}
\newcommand{\ptassoc} {\mbox{$p_{Tassoc}$}}
\newcommand{\kt} {\mbox{$k_T$}}

\newcommand{\vjt} {\mbox{$\vec{j}_{T}$}}

\newcommand{\jty} {\mbox{$j_{Ty}$}}
\newcommand{\mkt} {\mbox{$\langle k_{T}\rangle$}}
\newcommand{\mjt} {\mbox{$\langle j_{T}\rangle$}}
\newcommand{\mkty} {\mbox{$\langle|k_{Ty}|\rangle$}}
\newcommand{\mjty} {\mbox{$\langle|j_{Ty}|\rangle$}}
\newcommand{\mRMSkt} {\mbox{$\sqrt{\langle k^2_T\rangle}$ }}
\newcommand{\mz} {\mbox{$\langle z\rangle$}}
\newcommand{\mztrig} {\mbox{$\langle z_{trig}\rangle$}} 
 
\newcommand{\mpttrig} {\mbox{$\langle p_{Ttrig}\rangle$}} 
\newcommand{\mptassoc} {\mbox{$\langle p_{Tassoc}\rangle$}} 
\newcommand{\cf}{\mbox{\sl CF}}

\newcommand{\snn} {\mbox{$\sqrt{s_{NN}}$}}

\newcommand{\piz} {\mbox{$\pi^{0}$}}

\def\bgi{\begin{itemize}}
\def\endi{\end{itemize}}
\def\bge{\begin{equation}}
\def\ende{\end{equation}}
\def\bgc{\begin{center}}
\def\endc{\end{center}}
\def\gev{\mbox{~GeV\ }}
\def\gevc{\mbox{~GeV/$c$\ }}

\def\mevc{\mbox{~MeV/$c$\ }}

\def\la{\langle }
\def\ra{\rangle }

\def\Journal#1#2#3#4{{#1}{\bf #2}, #3 (#4)}

\def\NPA{{Nucl. Phys.}~{\bf A}}
\def\NPB{{Nucl. Phys.}~{\bf B}}
\def\PLB{{Phys. Lett.}~{\bf B}}

\def\PRL{Phys. Rev. Lett.\ }
\def\PRD{{Phys. Rev.}~{\bf D}}
\def\PRC{{Phys. Rev.}~{\bf C}}

\def\etall{{\it et al.}}

\begin{document}

\title[]{PHENIX measurement of jet properties and their modification
in heavy-ion collisions}

\author{Jan Rak\dag \hskip 2mm for the PHENIX Collaboration 
\footnote[3]{For the full PHENIX Collaboration list and acknowledgments, see Appendix 
``Collaborations'' of this volume}
}

\address{\dag\ Iowa State University, Ames, IA 50011-3160}

\begin{abstract}
The properties of jets produced in $p+p$, $d+Au$ and $Au+Au$
collisions at \snn=200~\gev\ are studied using the method of two
particle correlations. The trigger particle is assumed to be a leading
particle from a high \ptt\ jet while the associated particle is
assumed to come from either the same jet or the away jet. From the
angular width and yield of the same and away side correlation peaks,
the parameters characterizing the jet properties are extracted.
\end{abstract}



The high-\ptt\ particle yield measured at RHIC (\snn=130 and 200~GeV)
was found to be strongly suppressed in $Au+Au$ central collisions
\cite{RAA_02}. Furthermore, the measurement of high-\ptt\ particle yield
in $d+Au$ induced collisions \cite{RdAu} confirmed that the
suppression can be fully attributed to the final state interaction of
high-energy partons with an extremely opaque nuclear medium
\cite{quenching_WangMiklos,quenching_Wang}.
Parton propagation through the excited nuclear
medium should be accompanied by stimulated gluon radiation, which
results in the modification of fundamental properties of
hard-scattering such as broadening of parton transverse momentum \kt\
\cite{Urs_ktBroadening,Ivan_dAu} and modification of jet fragmentation
\cite{Wang_fragModif}. Thus the measurement of jet fragmentation
properties in heavy ion collisions, compared to the results for $p+p$
collisions, should provide more detailed insight into the process of
formation and materialization of excited nuclear medium.

\section{Jet angular correlations}

We explore the systematics of jet fragmentation by the method of
two-particle azimuthal correlations. This method, which worked well at
ISR energies ($\sqrt{s}$=63~GeV) and below \cite{CCORjt}, is an 
alternative method to the full jet reconstruction when the use of the
latter is difficult or impossible due to high-multiplicities, as in
heavy ion collisions.

\begin{figure}[h] 
\bgc
\resizebox{7.5cm}{5cm}{\includegraphics[viewport=0 0 565 500,clip]{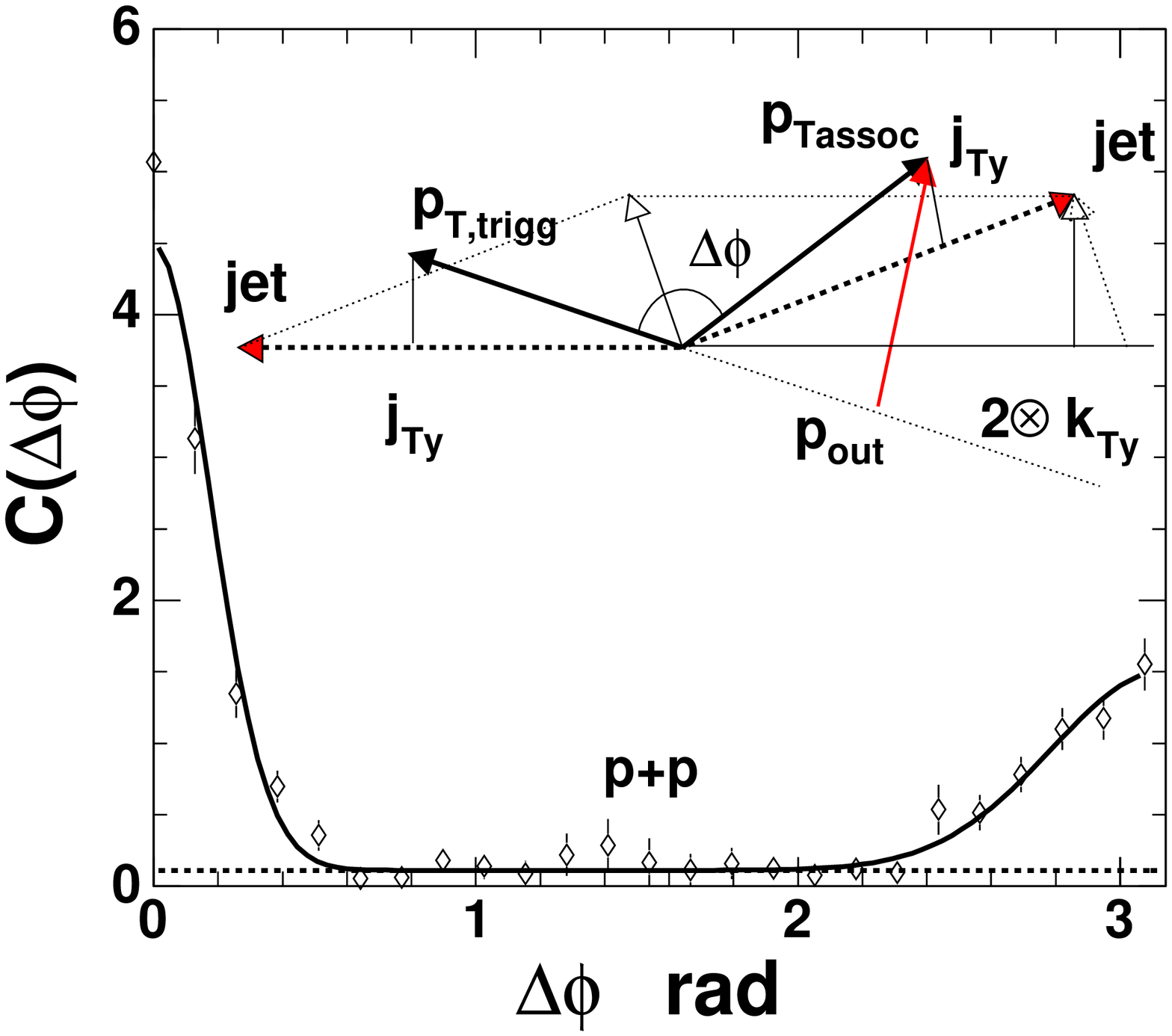}}
\resizebox{7.5cm}{5cm}{\includegraphics[viewport=0 0 565 500,clip]{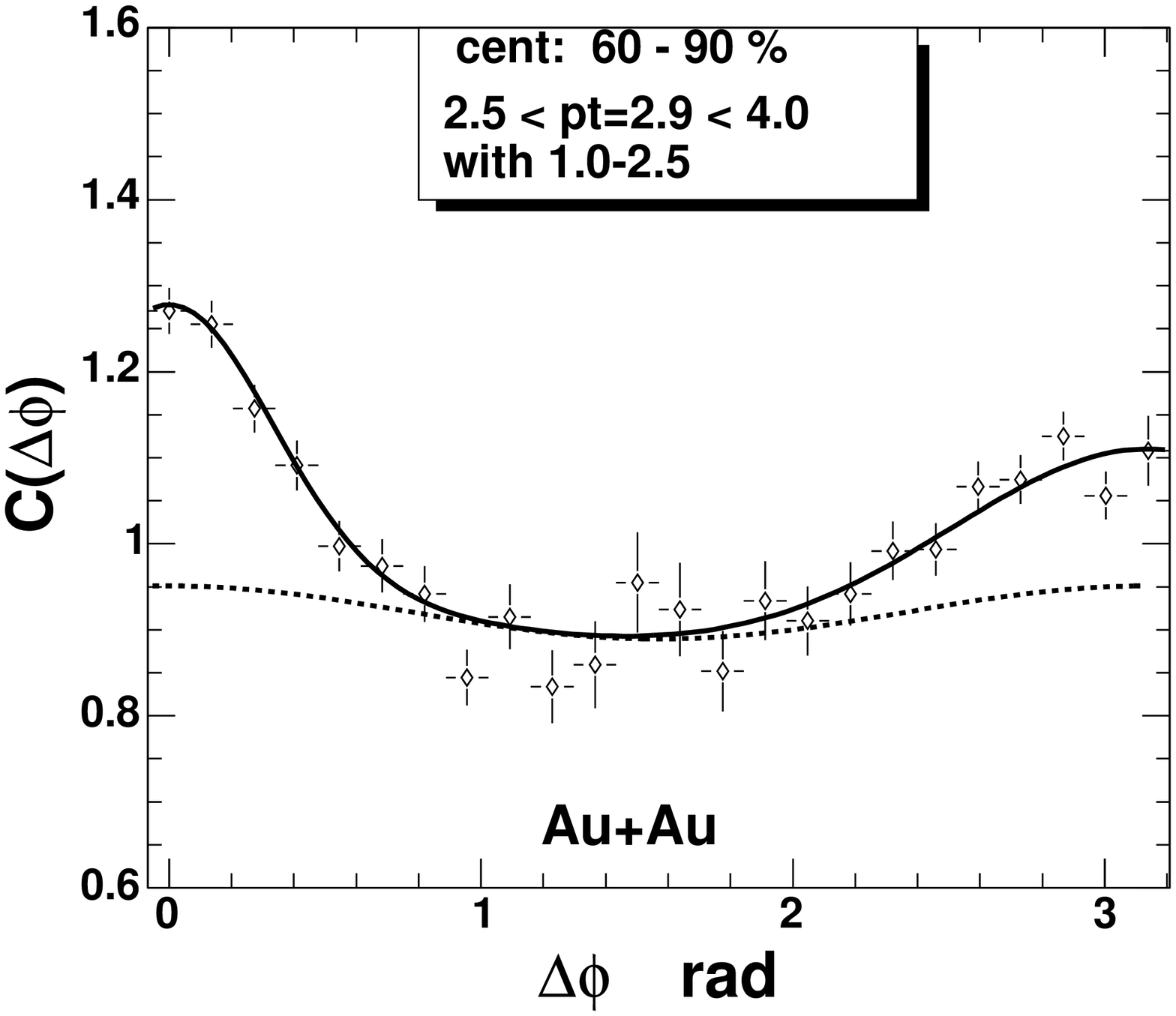}}
\caption{\label{CFnc}
{\sl Left:} 
Measured correlation functions in $p+p$ collisions for both 
particles in the $3.0<p_T<4.0$ range. Solid
line corresponds to the fit of the two Gaussian functions representing the
intra-jet and inter-jet correlation. Dashed line represents the
uncorrelated background distribution.
{\sl Right:}
The correlation functions from $Au+Au$ data with $2.5<\pttrig<4.0$ and $1.0<\ptassoc<2.5$.
The dashed line represent the $cos(2\Delta\Phi)$ background distribution and the solid
line represent the contribution of jet correlations.
}
\endc 
\end{figure}

Jets are produced by the hard scattering of two partons
\cite{Feynman1,Feynman4}. Two scattered partons propagate nearly 
back-to-back in azimuth from the collision point and fragment into the
jet-like spray of final state particles (see embedded schematics on
\fig{CFnc} where only one fragment of each parton is shown). These
particles have a momentum \vjt\ perpendicular to the partonic
transverse momentum, with component \jty\ projected onto the azimuthal
plane. The magnitude of \mjt\ measured at lower energies 
has been found to be $\sqrt{s}$ and \ptt\ independent \cite{CCORjt}. 

Despite the naive expectation of parton collinearity in the transverse
plane, it was found that each of the partons carry the effective
transverse momentum $\vec{k}_{T}$, originally described as
``intrinsic'' \cite{Feynman5}. The measurement of net transverse
momenta $\la p_{T}^2\ra_{pair} = 2\cdot\la k_{T}^2\ra$ of diphoton,
dimuon or dijet over the wide range of $\sqrt{s}$ in $p+p$ collisions
gives \mkt\ as large as 5\gevc\ \cite{kt_E609_Apana}. Clearly, this
value can not be attributed to the intrinsic transverse momentum
given by the constituent quark mass ($\approx$300~\mevc) and the NLO
re-summation techniques has to be invoked \cite{Werner}.
\vskip -0.5 cm
The analysis uses two-particle azimuthal correlation functions (\cf) to
measure the distribution of the azimuthal angle difference ($\Delta
\phi = \phi_1 - \phi_2$) between pairs of charged hadrons (see \fig{CFnc}).
The correlation function is defined as
$C(\Delta\phi)=\mathcal{N}\cdot{N_{cor}\over N_{uncor}}$ where
$N_{cor}$ and $N_{uncor}$ are the observed $\Delta\phi$ distributions
of charged particle pairs in the same or mixed events, respectively and
$\mathcal{N}$ is the normalization factor.

We fit the measured \cf\ by two gaussians, one for the near-side
component (around $\Delta\phi =0$) and one for the far-side component
(around $\Delta\phi =\pi$), and a constant for the uncorrelated pairs
in case of $p+p$ and $d+Au$ collisions. In case of $Au+Au$ correlation
functions, the background distribution has a ``harmonic'' form
$1+2a^2_2\cos{(2\Delta\phi)}$ with $a_2$ as a free parameter.

\begin{figure}[h]
\bgc
\resizebox{7.5cm}{5.5cm}{\includegraphics[viewport=0 0 514 500,clip]{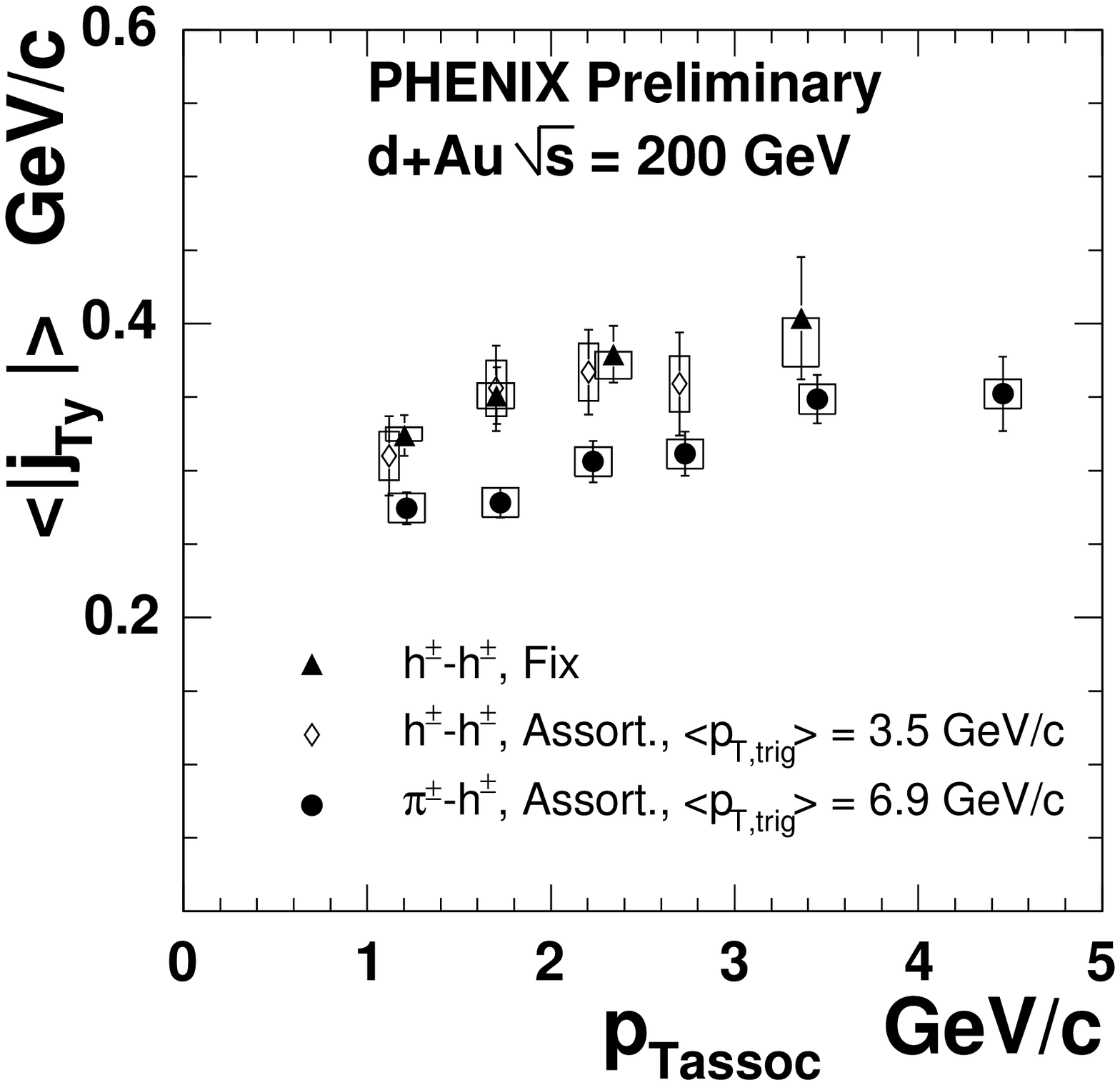}}
\resizebox{7.5cm}{5.5cm}{\includegraphics[viewport=0 0 514 500,clip]{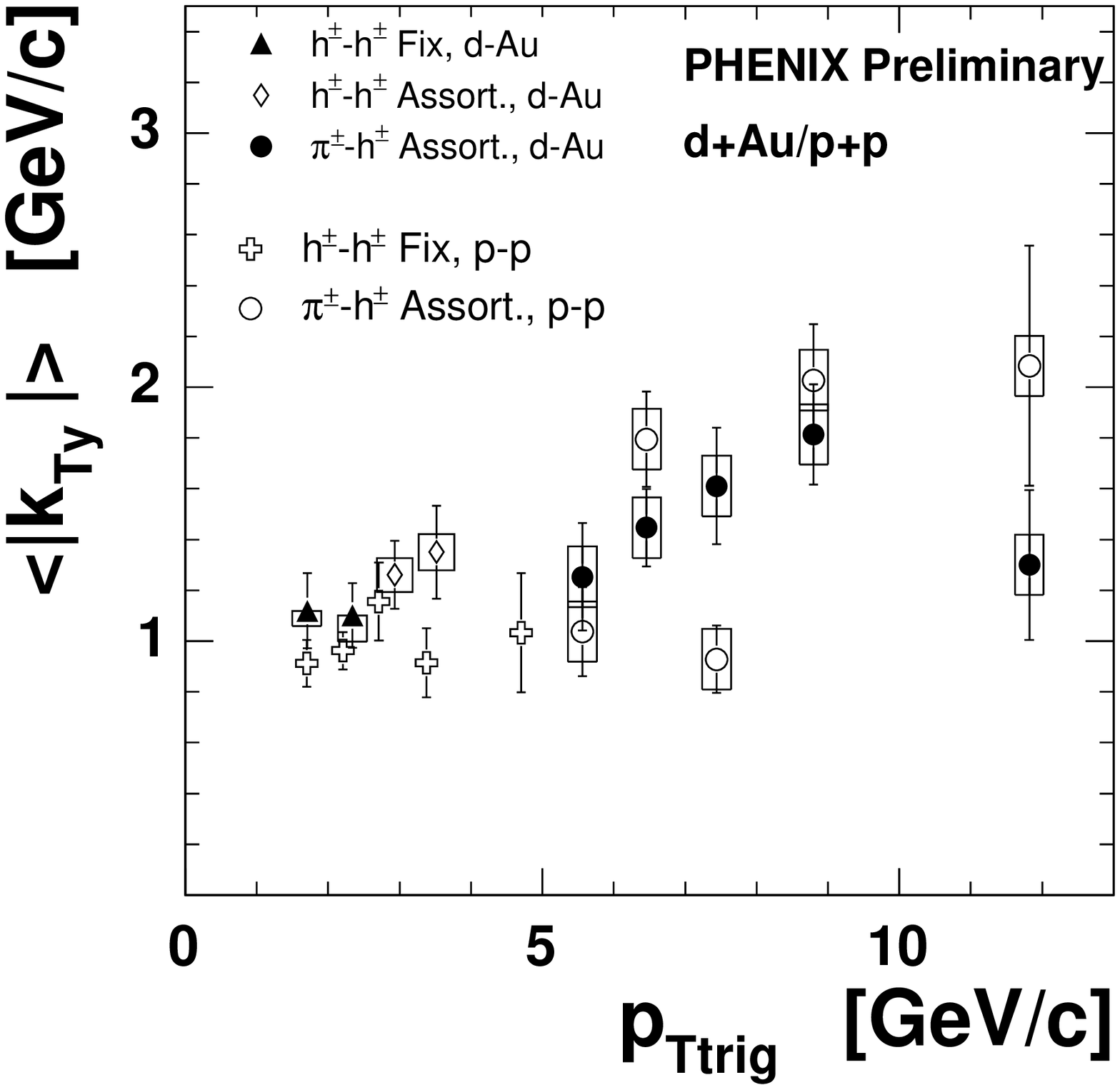}}
\caption{\label{jkt_ppdAu}
{\sl Left:}
Extracted values of \mjty\ as a function
of \ptassoc\ in $d+Au$.
{\sl Right:}
Variation of \mkty\ with \pttrig\ for $p+p$ and $d+Au$.
Boxes drawn around the data points indicate the systematic error bars.
}
\endc
\end{figure}

For two-particles with average transverse momenta \mpttrig\ and
\mptassoc\ from the same jet, the width of the near-side correlation, $\sigma_N$,
can be related to \mjty\ as
\bge
\langle| j_{T y} |\rangle \approx  \sqrt{2 \over \pi}
   {\mpttrig\mptassoc \over \sqrt{\mpttrig^2+\mptassoc^2}}\; \sigma_{N}
\label{eq_jt}
\ende
if we assume \mjty$\ll$\mpttrig\ and \mjty$\ll$\mptassoc.
In order to extract \mkt\ from the width of the away-side peak width
$\sigma_F$, we started with the relation between $\la|p_{out}|\ra$,
the average transverse momentum component of the away-side particle
$\vec{p}_T$ perpendicular to trigger particle $\vec{p}_{Ttrig}$ in the
azimuthal plane, and $k_{Ty}$ given in
\cite{CCORjt}, \cite{Feynman5}.  We note however, that
\cite{Feynman5} explicitly neglected $\mztrig=\la\pttrig/p_{Tjet}\ra$
in the formula at ISR energies, where $\mztrig\simeq$~0.85,
while it is not negligible at \snn=200~GeV. Taking $\mztrig$ into
account we derived
\bge\label{eq_kt}
\mkty\mztrig =
{1 \over x_h\sqrt{2}}
\sqrt{ \mptassoc^2\sin^2\sqrt{2\over\pi}\sigma_F - (1+x_h^2)\la|j_{Ty}|\ra^2}
\ende
where $x_h=\mptassoc/\mpttrig$. Equation (\ref{eq_kt}) indicates that in order
to extract the magnitude of \mkty\ the external knowledge of the
\mztrig\ is needed. We have analyzed the \ptt-dependence of the \piz\ invariant cross
section measured in $p+p$ collisions \cite{pi0ppPRL}. In the
$p_T>3$\gevc\ region where the power law function with the power of 
$n=8.05\pm0.06$ provides the best description of the data, we
assumed that the final state parton distribution function has the same
shape as the \piz\ invariant cross section and we found \mz$_{p_T>3}$=0.75$\pm$0.05.
Knowing \mztrig, $\sigma_N$ and $\sigma_F$ and using (\ref{eq_jt}), (\ref{eq_kt}) the
\mjty\ and \mkty\ can be calculated. The results for $p+p$ and $d+Au$
are shown on \fig{jkt_ppdAu}.

\vskip -0.5 cm
The \mjty\ and \mkty\ values are extracted using ``fixed'' and
``assorted''-\ptt\ correlation technique. The former method explores
the correlation between particle pairs of the similar values of
\pttrig\ and \ptassoc\, whereas the assorted method uses asymmetric
pairs. The values of \mjty\ in $p+p$ and $d+Au$ data are in good
agreement and the average value \mjty=324$\pm$6~MeV/c (\mjt=$\pi\over
2$\mjty=510$\pm$10~MeV/c) has been obtained.  Within the systematic
and statistical errors it is not clear whether or not the \mkt\ values
follow the rising trend seen at lower energies \cite{CCORjt}.
Nonetheless, averaging the $h^{\pm}-\pi^{\pm}$ and $h^{\pm}-h^{\pm}$
data over \pttrig\ in case of $p+p$ data we found 
\mkty$_{pp}$=1.08$\pm$0.05~GeV/c (\mRMSkt$_{pp}$=$\sqrt{\pi}$\mkty$_{pp}$=1.92$\pm$0.09~GeV/c) 
and for $d+Au$ data \mkty$_{dAu}$=1.36$\pm$0.07$\pm$0.12~GeV/c 
(\mRMSkt$_{dAu}$=2.42$\pm$0.12$\pm$0.21~GeV/c).

\begin{figure}[h]
\bgc
\resizebox{7.5cm}{5.5cm}{\includegraphics[viewport=0 0 565 500,clip]{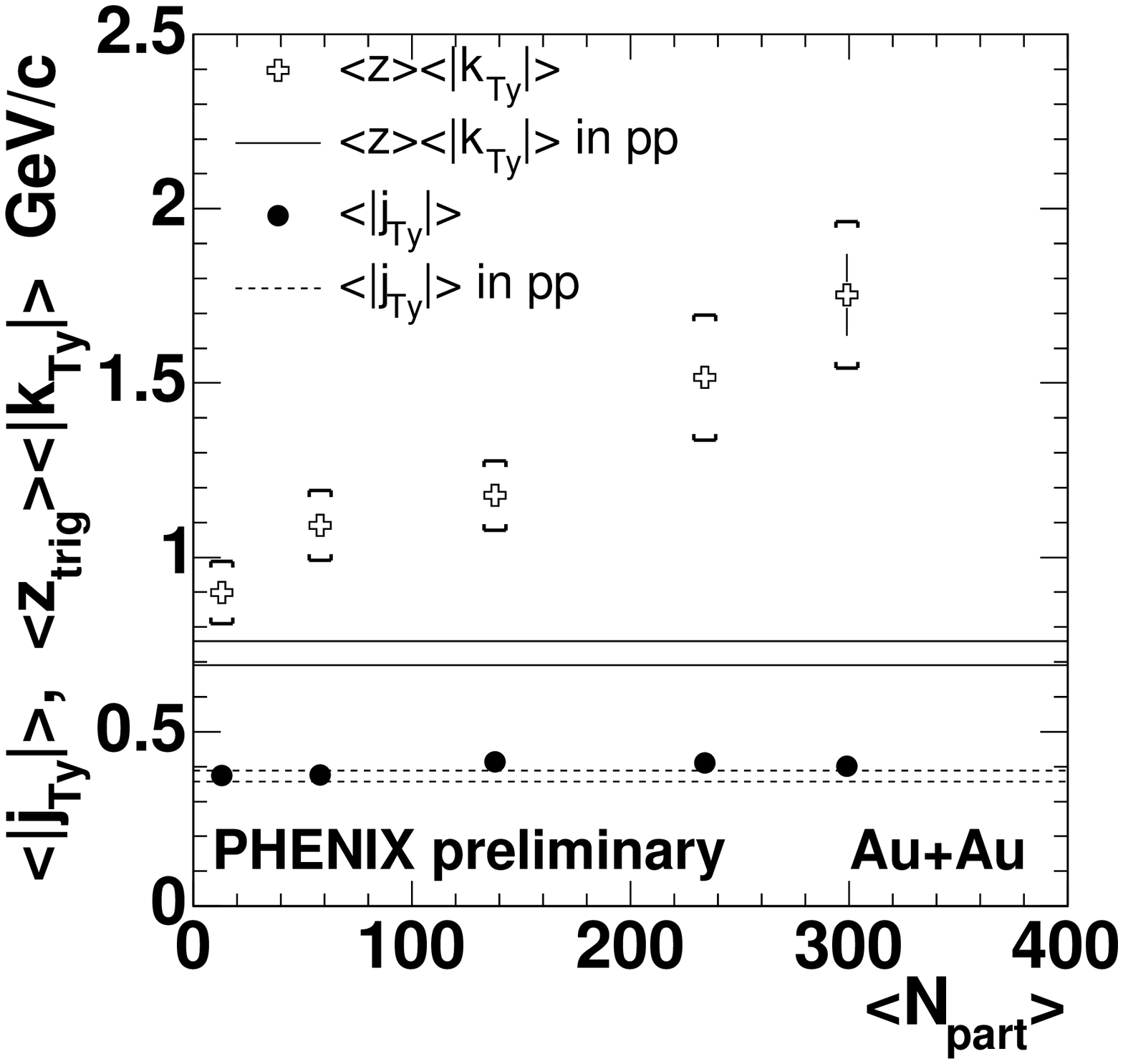}}
\resizebox{7.5cm}{5.5cm}{\includegraphics[viewport=0 0 565 500,clip]{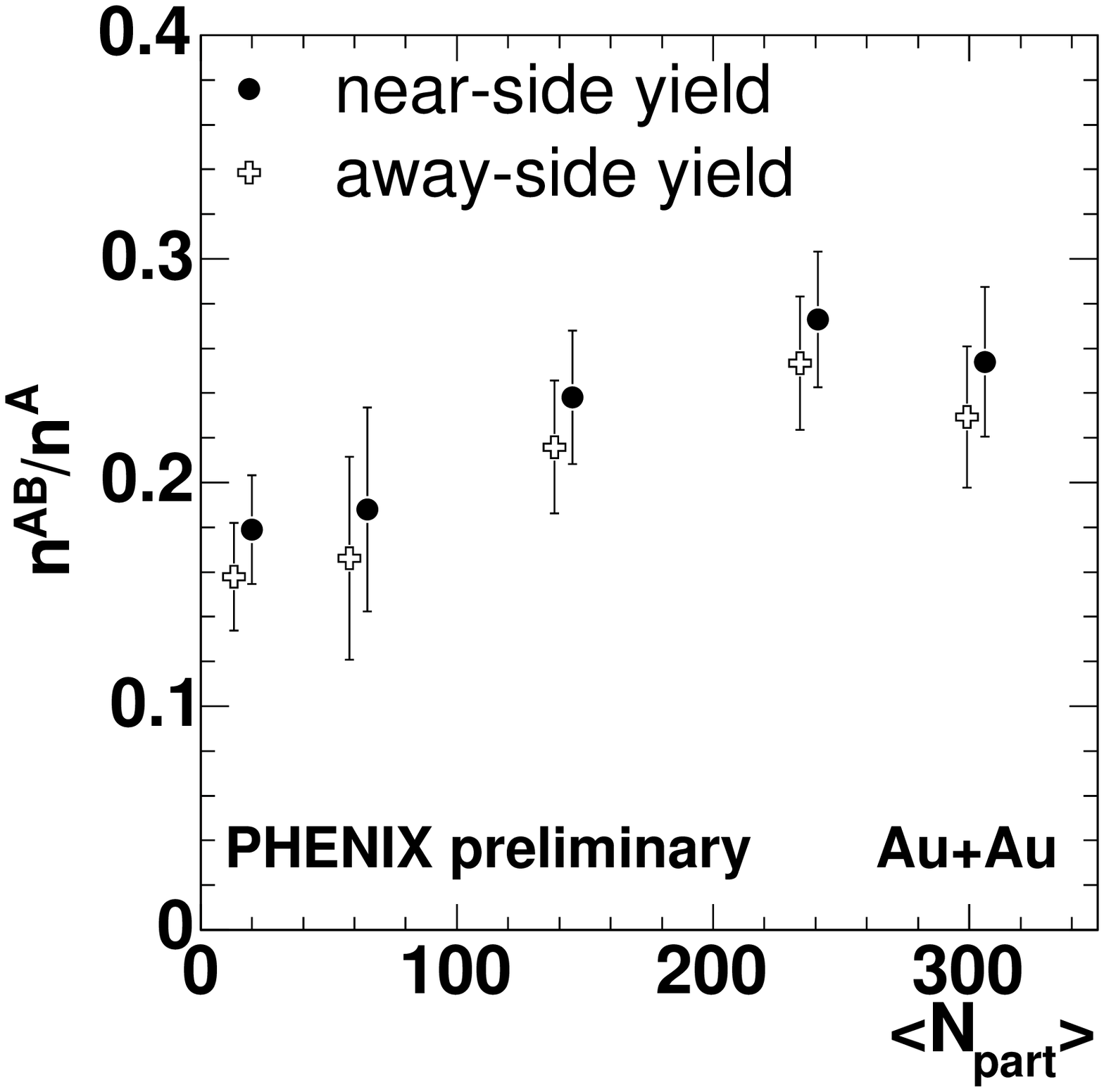}}
\caption{\label{jkt_Au}
{\sl Left:}
Centrality dependence of the \mjt\ and \mztrig\mkty in $Au+Au$ for
2.5$<$\pttrig$<$4.0\gevc\ and 1.0$<$\ptassoc$<$2.5\gevc\
The dashed and solid lines represent the values of \mjty\ and
\mztrig\mkty\ measured in $p+p$ data. Brackets indicate the systematic error bars.
{\sl Right:}
Associated near-side (solid circles)
and away-side (hollow crosses) yield of 1.0$<$\ptassoc$<$2.5\gevc\
particles per trigger particle found in the \ptt-bin
2.5$<$\pttrig$<$4.0\gevc. Errors are dominated by systematic error.  
}
\endc
\end{figure}

A similar analysis has been done also for $Au+Au$ data for various
collision centralities, characterized by the average number of
participants $\la N_{part}\ra$. We have studied the angular width and
the associated yield per trigger particle for
2.5$<$\pttrig$<$4.0\gevc\ and 1.0$<$\ptassoc$<$2.5\gevc.  However, in
this case, there is no justification of using $D(z)$ extracted from
$p+p$ data so we report only \mztrig\mkty\ (see \fig{jkt_Au}). Whereas
the \mjty\ values shows essentially no dependence on $N_{part}$, the
away-side width and \mztrig\mkty\ reveals a rather dramatic raising
trend compared to $p+p$ and $d+Au$ indicating strong final state partons' interaction with nuclear
medium.
\vskip -0.3 cm
\vskip -0.5 cm
The associated near and away-side yields in $Au+Au$ are found to be
rather constant or slightly rising with centrality in contrast of
measurement done at higher \pttrig-range
\cite{STAR_b2b_suppression}. This observation seems to indicate strong
jet rescattering rather than the full jet absorption in neclear
medium.


\vskip -0.5 cm
\section*{REFERENCES}
\begin{thebibliography}{100}

\bibitem{RAA_02}
	K.~Adcox \etall, \Journal{\PRL}{91}{072301}{2003}.

\bibitem{RdAu}
	S.~S.~Adler \etall, \Journal{\PRL}{91}{072303}{2003}.

\bibitem{quenching_WangMiklos} 
	X.~N.~Wang and M.~Gyulassy, \Journal{\PRL}{68}{1480}{1992}.

\bibitem{quenching_Wang}
	X.~N.~Wang, \Journal{\PRC}{58}{2321}{1998}.

\bibitem{Urs_ktBroadening}  
	C.~A.~Salgado and U.~A.~Wiedemann, hep-ph/0310079.

\bibitem{Ivan_dAu}
	J.~Qiu, I.~Vitev, nucl-th/0306039.

\bibitem{Wang_fragModif}  
	X.~N.~Wang, \Journal{\NPA}{702}{238}{2002}; X.~N.~Wang, nucl-th/0305010.

\bibitem{CCORjt} A.L.S. Angelis \etall, \Journal{\PLB}{97}{163}{1980}.

\bibitem{Feynman1}
	S.~M.~Berman, J.~D.~Bjorken, J.~B.~Kogut, \Journal{\PRD}{4}{3388}{1971}.
\bibitem{Feynman4}
	J.~F.~Owens and J.~D.~Kimel, \Journal{\PRD}{18}{3313}{1978}.

\bibitem{Feynman5}
	R.~P.~Feynman, R.~D.~Field and G.~C.~Fox, \Journal{\NPB}{128}{1}{1977}.

\bibitem{kt_E609_Apana}
	L.~Apanasevich \etall, \Journal{\PRD}{59}{}{1999}.

\bibitem{Werner}
	D.~Boer, W.~Vogelsang, hep-ph/0312320.

\bibitem{pi0ppPRL}
	S.~S.~Adler \etall, submitted to Phys. Rev. Lett., hep-ex/0304038.


\bibitem{STAR_b2b_suppression}
        C.~Adler \etall, \Journal{\PRL}{90}{082302}{2003}.

\end{thebibliography}
\end{document}